\documentclass{llncs}
\usepackage{graphicx}
\usepackage{multirow}
\usepackage{amsmath}

\spnewtheorem{Nota}{Notation}{\bfseries}{\itshape}
\spnewtheorem{Obs}{Observation}{\bfseries}{\itshape}
\spnewtheorem{Theo}{Theorem}{\bfseries}{\itshape}

  \newcommand{\BserI}{\overleftarrow{\scriptscriptstyle BseRI}}
  \newcommand{\FokI}{\overrightarrow{\scriptscriptstyle FokI}}
  \newcommand{\BsrDI}{\overrightarrow{\scriptscriptstyle BsrDI}}
  \newcommand{\BpmII}{\overleftarrow{\scriptscriptstyle BpmI}}
  \newcommand{\BpmID}{\overrightarrow{\scriptscriptstyle BpmI}}
  \newcommand{\BbvI}{\overleftarrow{\scriptscriptstyle BbvI}}

\begin{document}
	\title{Logical N-AND Gate on a Molecular Turing Machine}
	\author{Victor Hernandez-Urbina\inst{1}}
	\institute{Institute of Perception, Action and Behavior, University of Edinburgh. \email{j.v.hernandez-urbina@ed.ac.uk}}
	
	\maketitle
	\begin{abstract}
		In Boolean algebra, it is known that the logical function that corresponds to the negation 
		of the conjunction --NAND-- is universal in the sense that any other 
		logical function can be built based on it. 
		This property makes it essential to modern digital electronics and computer processor design.
		Here, we design a molecular Turing machine that 
		computes the NAND function over binary strings of arbitrary length. 
		For this purpose, we will perform a mathematical abstraction of
		the kind of operations that can be done over a double-stranded DNA molecule, as well as presenting
		a molecular encoding of the input symbols for such a machine.
		\keywords{molecular computing, DNA computing, Turing machine, Boolean functions, restriction enzymes.}
	\end{abstract}
	\section{Introduction}
		Moore's Law is the empirical observation that the number of transistors on integrated circuits doubles approximately every two years.
		However, it is known that the continuous miniaturization of electronic devices has undesirable consequences: fitting more and more chips
		in an area that is becoming smaller according to the standards of the market results in an uncontrollable increase of heat when the device
		begins its operation \cite{moore1965cramming}. 
		Hardware manufacturers are reaching the physical limit when attempting to develop more powerful devices in the the least space possible.
		Moreover, the holy grail of digital storage is being able to store a bit per atom, which implies massive storage in a very small volume. However, theory says
		that in order to achieve such an endeavour, one would have to pay the quantum price for dealing with subatomic particles.
		
		An alternative to electronic and quantum computing is \emph{molecular computing}, which also attempts to reduce the scale of computation and at the same time
		increase its power. Molecular computing uses as hardware (rather \emph{wetware}) organic molecules
		such as ribonucleic acid (RNA) and deoxyribonucleic acid (DNA). In this work, we focus on the latter, which is known elsewhere as DNA computing.
		
		There are two main reasons that make DNA an attractive computing medium, one is \emph{parallelism} and the other, \emph{complementarity}.
		The density of information encoded in a DNA strand, plus its relatively ease to generate copies of it (via techniques such as PCR \cite{bartlett2003short}), 
		offers the possibility of realising parallel computing massively. 
		For example, Leonard Adleman, one of the pioneers of this computing paradigm, presented in \cite{adleman1998computing} a solution to the 
		\emph{Hamiltonian Path Problem} by probing in parallel different solutions to this NP-complete problem.
		On the other hand, complementarity refers to the fact that a strand of DNA is composed by two single strands of nucleotides that \emph{bind naturally} 
		through their bases. 
		There are four of such bases: guanine (G), adenine (A), thymine (T), and cytosine (C). Given their chemical composition, nucleotides with base G bind
		naturally to nucleotides with base C, this fact makes them complementary bases, whereas nucleotides with base A bind to those with base T.
		As mentioned before, this complementarity is given free by nature and it can be exploited to model and encode a particular task.
		
		Representing a double strand of DNA as two single strands, one above the other taking into account the complementarity of their bases,
		is a big simplification when actually both strands are coupled forming the well-known double helix.
		Our representation of such strings in terms of a strand that extends itself horizontally will help us through our exposition in the following pages.
		However, we should also mention that in this work we adopt the convention of expressing the direction of a single string of DNA as going from the $5'-$end to the $3'-$end, 
		which is given by the nature of its chemical components and that is used extensively in biochemistry \cite{paun1998dna}.
		
		DNA is one of the main computing media in molecular computation, and it is crucial in the processes inside the living cells of organisms.
		Its two main functions are the codification for the creation of new proteins, and self-replication, such that an exact copy of itself is inherited to
		cellular offspring. As a computing medium it has been used to solve the Hamiltonian Path Problem and other combinatorial problems
		\cite{adleman1994molecular}, as well as being used as a mechanism to detect the presence of molecules that signal the emergence of a chronic disease
		\cite{benenson2004autonomous}\cite{shapiro2006bringing}.
		In \cite{farfan2010applying} we present the design of an enzyme-free molecular machine that detects the emergence of hepatic fibrosis, and upon detection
		releases a molecule to interfere with the over expressed gene.
				
		In this paper, our motivation comes from the work of Vineet Gupta et al. \cite{gupta1997arithmetic} where the authors present a technique to
		simulate logic and arithmetic functions through DNA molecules.
		However, the issue that arises in such work is that the method suggested lacks proper automation, that is, depends greatly in the presence
		of human operation, which contradicts the essence of computation, namely, the automation of processes that would imply a big effort, cost, and
		time in a situation in which those resources are limited.
		Our suggestion, thus, is the design of a DNA machine, which mimics the behaviour a particular logical function autonomously, that is,
		minimising the intervention of an external operator. The logic function that we will simulate molecularly is the inverted AND gate, also known as
		NAND gate, which is functionally complete. This fact implies that any other logic function can be expressed as a combination of gates of this
		type. Hence, its importance in the design of digital electronics, such as computer processors.
		
		We make use of a set of operations that can be performed on a DNA strand. This set can be regarded as a gift from biochemistry to
		this novel computing paradigm. The kind of operations that can be performed on a DNA strand are: duplication, concatenation, cleavage, extension, length measurement,
		among others. For the type of problem that we address here, we need only a small set of operations, which we briefly describe below. 
		For a more exhaustive description we refer the reader to \cite{paun1998dna}.
		
		\begin{itemize}
			\item\emph{Measuring a DNA strand's length.} The length of a double strand of DNA is given by the number of base pairs (a nucleotide and 
			its complement) and is denoted by the unit \emph{bps}. To measure this length a technique known as gel electrophoresis is employed.
			\item\emph{Concatenation and separation of a DNA strand.} DNA strands are joined by a covalent bond horizontally (that is, when nucleotides
			are adjacent in a single strand), and by a hydrogen bond vertically (that is, when bases are complementary). The latter is weaker than the former,
			so it is possible to disjoin two complementary strings of DNA leaving the nucleotides in each single string intact just by increasing temperature.
			If the solution is cooled down again, the single strings are susceptible of joining again. This process must be performed slowly so strings have
			a chance to recognise their complementary bases. 
			Then, a type of enzyme known as ligase is used, which joins covalently
			two consecutive nucleotides \cite{paun1998dna}.
			\item\emph{Cutting a DNA strand.} A double-stranded DNA molecule can be cleaved by the action of enzymes. Their action can be performed
			at the ends of a molecule or within it. The latter type of enzymes is known as endonucleases, or restriction enzymes.
			These are found in bacteria and they provide their hosts with a defense mechanism against viruses.
			These enzymes acquire their name from the organism from which they were obtained. For example, enzyme EcoRI receives its name because it
			was the first enzyme to be identified (hence, \emph{I}) in the bacteria \emph{E. coli} strain R.
			Restriction enzymes work by identifying a particular DNA string and then cleaving the molecule either at this recognised site or a few nucleotides away
			from it. In this work, we use this latter type of enzymes, which are known as restriction enzymes type IIS \cite{pingoud2001structure}.
		\end{itemize}
		
		With these ideas in mind, we present our model in the following pages. 
		The structure of this paper is as follows, in the next section we present the mathematical
		version of the Turing machine that computes the logical NAND function. 
		In section \ref{sec:MTM_2}, we present the molecular version of this Turing machine, and we present the main result of our work, which is in the shape of a
		theorem summarising the mathematical abstraction of the elements comprising the molecular machine.
		Finally, in section \ref{sec:CONC} we discuss some of the implications of our model along with some conclusions.
		
\section{Model}
	
		Broadly speaking, a Turing Machine (TM) is a mathematical object that mimics the action of a machine consisting of a \emph{head} that reads and writes on a  tape
		obeying a fixed set of rules. 
		Although simple in its design, a Turing machine is a powerful computing device as it can be adapted to mimic the logic of any computer algorithm.
		
		In theory, the tape of the machine has infinite length and its head moves one square at a time along it according to a program specified
		as a set of rules. Formally, a TM is defined as follows:
		
		\begin{definition}
			A Turing Machine is the $7-$tuple $(S, \Sigma, \beta, \Sigma^{*}, T, s_{0}, A)$, which consists of:
			\begin{enumerate}
				\item A finite set of states $S$.
				\item A finite set of symbols (or alphabet) $\Sigma$.
				\item A symbol $\beta\in\Sigma$ representing a blank character.
				\item A finite set $\Sigma^{*} \subset\Sigma\setminus\{\beta\}$ of input symbols.
				\item A transition function $T: S\times\Sigma\to S\times\Sigma\times\{ L,R\}$, where $L$ ($R$) denotes a single left (right) movement
				of the head on the tape.
				\item An initial state $s_{0}\in S$.
				\item A finite set of accepting states $A\subset S$, for which the machine halts and the computation is over.
			\end{enumerate}
		\end{definition}
		
		As mentioned before, the logic function that we implement on a TM is an inverted AND gate, whose truth table is shown in table \ref{table1}.
		Our TM reads two binary strings of length $n$ from its tape, and then computes the NAND function entry by entry following the rules specified by 
		its truth table.
		As a result, the machine writes a binary string of length $n$ on its tape with the result of the computation.
		
		\begin{table}[!h]
			\caption{NAND truth table}
			\label{table1}
			\begin{center}
				\begin{tabular}{|c|c|c|}
					\hline
					p & q & p NAND q \\
					\hline
					0 & 0 & 1 \\
					0 & 1 & 1 \\
					1 & 0 & 1 \\
					1 & 1 & 0 \\
					\hline
				\end{tabular}
			\end{center}
		\end{table}
		
		The first convention that we will adopt concerns the way in which we write the input strings on the machine's tape. 
		Let $A = (a_{1}, a_{2},\ldots,a_{n})$ and $B = (b_{1}, b_{2},\ldots,b_{n})$ be two binary strings of length $n$.
		We write the strings $A$ and $B$ intercalating each element of their entries on the tape. That is:
		
		\begin{displaymath}
			\label{desc:CintaTM}
			\begin{tabular}{c|c|c|c|c|c|c|c|c}
				\hline
				\multirow{2}{*}{$\cdots$} & \multirow{2}{*}{$a_{1}$} & \multirow{2}{*}{$b_{1}$} & \multirow{2}{*}{$a_{2}$} &
				\multirow{2}{*}{$b_{2}$} & \multirow{2}{*}{$\cdots$} & \multirow{2}{*}{$a_{n}$} & \multirow{2}{*}{$b_{n}$}
				& \multirow{2}{*}{$\cdots$}\\
				& & & & & & & & \\
			\hline
			\end{tabular}
		\end{displaymath}	
		
		With this in mind, we define our TM in the following way. Our alphabet is the set $\Sigma = \{0, 1, \varepsilon, \beta\}$, where $0$ and $1$ are the
		characters comprising our binary input and output strings, $\varepsilon$ is a symbol that will serve as an error detection mechanism at the
		end of the computation, and $\beta$ is our blank character. Thus, $\Sigma^{*} = \{0,1\}$.
		We consider the set of states $S = \{S_{0},S_{1},S_{2},HALT\}$, where $HALT\in A$ represents the accepting state which stops the computation,
		and $S_{0}$ is the initial state of the machine.
		Finally, the transition rules of our machine, and which ultimately mimic the behaviour of a NAND gate, are the following:
		\begin{equation}
			\label{desc:t1TM}
			\langle S_{0}, 0\rangle\mapsto\langle S_{1},\beta,\Rightarrow\rangle
		\end{equation}
		\begin{equation}
			\label{desc:t2TM}
			\langle S_{0}, 1\rangle\mapsto \langle S_{2},\beta,\Rightarrow\rangle
		\end{equation}
		\begin{equation}
			\label{desc:t3TM}
			\langle S_{0}, \beta\rangle\mapsto \langle HALT\rangle
		\end{equation}
		\begin{equation}
			\label{desc:t4TM}
			\langle S_{1}, 0\rangle\mapsto \langle S_{0},1,\Rightarrow\rangle
		\end{equation}
		\begin{equation}
			\label{desc:t5TM}
			\langle S_{1}, 1\rangle\mapsto \langle S_{0},1,\Rightarrow\rangle
		\end{equation}
		\begin{equation}
			\label{desc:t6TM}
			\langle S_{1}, \beta\rangle\mapsto \langle S_{0},\epsilon,\Rightarrow\rangle
		\end{equation}
		\begin{equation}
			\label{desc:t7TM}
			\langle S_{2}, 0\rangle\mapsto \langle S_{0},1,\Rightarrow\rangle
		\end{equation}
		\begin{equation}
			\label{desc:t8TM}
			\langle S_{2}, 1\rangle\mapsto \langle S_{0},0,\Rightarrow\rangle
		\end{equation}
		\begin{equation}
			\label{desc:t9TM}
			\langle S_{2}, \beta\rangle\mapsto \langle S_{0},\varepsilon,\Rightarrow\rangle
		\end{equation}
	
		Each of the transitions are interpreted in the following way: 
		\begin{displaymath}
			\langle current\_state, character\_being\_read\rangle\mapsto \langle new\_state,character\_to\_write,move\_head\_to\rangle
		\end{displaymath}
		
		In the following section we will implement all the elements of this TM into DNA molecules and we will provide a mathematical proof of its
		operation.
	\subsection{Molecular Turing Machine}
		\label{sec:MTM_2}
		In this section we present the molecular version of the TM defined previously. We perform a molecular encoding of all the elements of
		the TM recently described, which means that we will represent the alphabet, the transitions and states as DNA strands.
		In summary, a circular double-stranded DNA molecule of DNA will act as the tape of the TM, in which a substring will simulate the action of the 
		head.		
		Using ideas from \cite{rothemund1996dna}, we use restriction enzymes to design the hardware of our machine.
		As mentioned previously, we use type IIS restriction enzymes to cut DNA strands at a defined distance from their non-palindromic asymmetric recognition sites
		\cite{pingoud2001structure}, which in other words means that the recognition and the cleavage sites are away from each other.
		
		In table \ref{table2} we describe the five restriction enzymes that we employ, their recognition site, the distance between it and the place the cut is performed, and the 
		direction in which they perform such cut. The nucleotide $N_{i}$ denotes an arbitrary base.
		We must point out that these enzymes cut from right to left, or vice versa, whenever they find their recognition sites in the top string (in direction $5'$ to $3'$)
		or its mirrored version in the bottom string (in direction $3'$ to $5'$). 
		However, we only specify the direction of cleavage in the way we require for the purposes of our machine.
		After cleavage these enzymes leave a sticky-end in the strings that have been cut. This sticky-end is an overhang of nucleotides in which another
		molecule can be joined.
		We group these five enzymes in a mathematical set that we name $\Phi$.
				
		\begin{table}[!h]
			\caption{Restriction Enzymes}
			\label{table2}
			\begin{center}
				\begin{tabular}{|c|c|c|c|}
					\hline
					Name & Direction & Recognition Site & Restriction Site\\
					\hline
					\hline
					FokI & $\rightarrow$ &
					\begin{tabular}{c}
						$5'-$GGATG$-3'$\\
						$3'-$CCTAC$-5'$\\
    					\end{tabular} & 
					\begin{tabular}{l}
						$5'-\ldots N_{9}$\\
						$3'-\ldots N_{9}N_{10}N_{11}N_{12}N_{13}$\\
    					\end{tabular} \\
					\hline
					BsrDI & $\rightarrow$ &
					\begin{tabular}{c}
						$5'-$GCAATG$-3'$\\
						$3'-$CGTTAC$-5'$\\
    					\end{tabular} & 
					\begin{tabular}{l}
						$5'-N_{1}N_{2}$\\
						$3'-\ldots $\\
    					\end{tabular} \\
					\hline
					BpmI & $\rightarrow$ &
					\begin{tabular}{c}
						$5'-$CTGGAG$-3'$\\
						$3'-$GACCTC$-5'$\\
    					\end{tabular} & 
					\begin{tabular}{l}
						$5'-\ldots N_{14}N_{15}N_{16}$\\
						$3'-\ldots N_{14}$\\
    					\end{tabular} \\
					\hline
					BpmI & $\leftarrow$ &
					\begin{tabular}{c}
						$5'-$CTCCAG$-3'$\\
						$3'-$GAGGTC$-5'$\\
    					\end{tabular} & 
					\begin{tabular}{r}
						$N_{14}\ldots-3'$\\
						$N_{16}N_{15}N_{14}\ldots-5'$\\
    					\end{tabular} \\
					\hline
					BserI & $\leftarrow$ &
					\begin{tabular}{c}
						$5'-$CTCCTC$-3'$\\
						$3'-$GAGGAG$-5'$\\
    					\end{tabular} & 
					\begin{tabular}{r}
						$N_{8}\ldots-3'$\\
						$N_{10}N_{9}N_{8}\ldots-5'$\\
    					\end{tabular} \\
					\hline
					BbvI & $\leftarrow$ &
					\begin{tabular}{c}
						$5'-$GCTGC$-3'$\\
						$3'-$CGACG$-5'$\\
    					\end{tabular} & 
					\begin{tabular}{r}
						$N_{12}N_{11}N_{10}N_{9}N_{8}\ldots-3'$\\
						$N_{8}\ldots-5'$\\
    					\end{tabular} \\
					\hline
				\end{tabular}
			\end{center}
		\end{table}		 
		
		To simplify our exposition we adopt the following conventions regarding the representation of DNA strands, in particular, the representation
		of the recognition sites and cleavage direction of the restriction enzymes.

		\begin{Nota}
			A DNA strand (that is not linked by its extremes) will be represented as a string of characters enclosed by square brackets.
			\begin{displaymath}
				\left[~~~~~~~~\right]
			\end{displaymath}
			Whereas, a DNA strand that is linked by its extremes will be represented as a string of characters enclosed by simple parentheses.
			\begin{displaymath}
				\left(~~~~~~~~\right)
			\end{displaymath}
		\end{Nota}

		\begin{Nota}
			The string $\left[\overrightarrow{RS}\right]$ denotes the recognition site of restriction enzyme $RS\in\Phi$ 
			whose cleavage direction is from left to right.
			Analogously, the string $\left[\overleftarrow{RS}\right]$ denotes the same enzyme, but this time its cleavage is from right to left.
		\end{Nota}
				
		We will use a word size of $6$ bps to encode the symbols in the set $\Sigma$, plus a suffix of $4$ bps, which serves as character delimiter
		in the molecular tape.
		We do not give an explicit declaration of the bases comprising the strands from the alphabet; as long as they do not contain any
		of the recognition sites used in the set $\Phi$, the choice of their bases is arbitrary. Thus, our symbols, plus the suffix, look like:
		
		\begin{equation}
			0 = \left[ \frac{\scriptstyle A_{1}A_{2}A_{3}A_{4}A_{5}A_{6}}{\scriptstyle \widetilde{A_{1}}\widetilde{A_{2}}
			\widetilde{A_{3}}\widetilde{A_{4}}\widetilde{A_{5}}\widetilde{A_{6}}}|\frac{\scriptstyle F_{1}F_{2}F_{3}F_{4}}
			{\scriptstyle \widetilde{F_{1}}\widetilde{F_{2}}\widetilde{F_{3}}\widetilde{F_{4}}}\right]
		\end{equation}		
		
		\begin{equation}
			\label{symb1}
			1 = \left[ \frac{\scriptstyle B_{1}B_{2}B_{3}B_{4}B_{5}B_{6}}{\scriptstyle \widetilde{B_{1}}\widetilde{B_{2}}
			\widetilde{B_{3}}\widetilde{B_{4}}\widetilde{B_{5}}\widetilde{B_{6}}}|\frac{\scriptstyle F_{1}F_{2}F_{3}F_{4}}
			{\scriptstyle \widetilde{F_{1}}\widetilde{F_{2}}\widetilde{F_{3}}\widetilde{F_{4}}}\right]
		\end{equation}

		\begin{equation}
			\beta = \left[ \frac{\scriptstyle \beta_{1}\beta_{2}\beta_{3}\beta_{4}\beta_{5}\beta_{6}}{\scriptstyle \widetilde{\beta_{1}}\widetilde{\beta_{2}}
			\widetilde{\beta_{3}}\widetilde{\beta_{4}}\widetilde{\beta_{5}}\widetilde{\beta_{6}}}|\frac{\scriptstyle F_{1}F_{2}F_{3}F_{4}}
			{\scriptstyle \widetilde{F_{1}}\widetilde{F_{2}}\widetilde{F_{3}}\widetilde{F_{4}}}\right]
		\end{equation}

		\begin{equation}
			\varepsilon = \left[ \frac{\textstyle \varepsilon_{1}\varepsilon_{2}\varepsilon_{3}\varepsilon_{4}\varepsilon_{5}\varepsilon_{6}}{\textstyle				\widetilde{\varepsilon_{1}}\widetilde{\varepsilon_{2}}
			\widetilde{\varepsilon_{3}}\widetilde{\varepsilon_{4}}\widetilde{\varepsilon_{5}}\widetilde{\varepsilon_{6}}}|\frac{\scriptstyle F_{1}F_{2}F_{3}F_{4}}
			{\scriptstyle \widetilde{F_{1}}\widetilde{F_{2}}\widetilde{F_{3}}\widetilde{F_{4}}}\right]
		\end{equation}
		
		Moreover, during computation time the current state of the TM is decoded in the sticky-end of the last cleaved molecule.
		Let us consider the following DNA strand:
		\begin{displaymath}
			N = \left[ \frac{\scriptstyle N_{1}N_{2}N_{3}N_{4}N_{5}N_{6}}{\scriptstyle \widetilde{N_{1}}\widetilde{N_{2}}
			\widetilde{N_{3}}\widetilde{N_{4}}\widetilde{N_{5}}\widetilde{N_{6}}}|\frac{\scriptstyle F_{1}F_{2}F_{3}F_{4}}
			{\scriptstyle \widetilde{F_{1}}\widetilde{F_{2}}\widetilde{F_{3}}\widetilde{F_{4}}}\right]
		\end{displaymath}
		
		\begin{enumerate}
			\item The following situation reports that the TM is in state $S_{0}$ with input $N$.
				\begin{displaymath}
					N = \left[ \frac{\scriptstyle
					N_{1}N_{2}N_{3}N_{4}N_{5}N_{6}}{\scriptstyle ~~~~~~~~~~~~
					\widetilde{N_{5}}\widetilde{N_{6}}}|\frac{\scriptstyle F_{1}F_{2}F_{3}F_{4}}
					{\scriptstyle \widetilde{F_{1}}\widetilde{F_{2}}\widetilde{F_{3}}\widetilde{F_{4}}}\right]
				\end{displaymath}
			\item The following situation reports that the TM is in state $S_{1}$ with input $N$.
				  \begin{displaymath}
					N = \left[ \frac{\scriptstyle
					N_{1}N_{2}N_{3}N_{4}N_{5}N_{6}}{\scriptstyle ~~~~~~~~~~~~~~~
					\widetilde{N_{6}}}|\frac{\scriptstyle F_{1}F_{2}F_{3}F_{4}}
					{\scriptstyle \widetilde{F_{1}}\widetilde{F_{2}}\widetilde{F_{3}}\widetilde{F_{4}}}\right]
				\end{displaymath}
			\item The following situation reports that the TM is in state $S_{2}$ with input $N$.
				\begin{displaymath}
					N = \left[ \frac{\scriptstyle
					N_{1}N_{2}N_{3}N_{4}N_{5}N_{6}}{}|\frac{\scriptstyle F_{1}F_{2}F_{3}F_{4}}
					{\scriptstyle \widetilde{F_{1}}\widetilde{F_{2}}\widetilde{F_{3}}\widetilde{F_{4}}}\right]
				\end{displaymath}
		\end{enumerate}
		
		We further extend our notation in order to make this exposition clearer.
		\begin{Nota}
			The string $\left[ X_{Y}\right]$
			where $X$ is a natural number, and $Y\in\Sigma$, represents $X$ number of nucleotides corresponding to symbol $Y$.
			Whereas the string $\left[ X\right]$
			represents $X$ number of arbitrary nucleotides.
		\end{Nota}		
		
		For example, the string $\left[ 6_{1}|4_{F}\right]$ is the simplified version of the DNA strand (\ref{symb1}) described above.
		With this in mind we are in position to present the molecular version of the transition rules that were described in the previous section. 
		\begin{equation}
			\label{mol:T1}
			\begin{gathered}
				T_{1}: <S_{0}, 0>\mapsto <S_{1},\beta,\Rightarrow> \\
				\left[ \overrightarrow{BsrDI}|4_{F}|6_{\beta}|
				4_{F}|6|\overleftarrow{BserI}|\overrightarrow{FokI}|4|4_{F}|12|
				\overleftarrow{BpmI}|\overrightarrow{BpmI}|8|6_{0}|6|\overleftarrow{BbvI}
				\right]
			\end{gathered}
		\end{equation}
		\begin{equation}
			\label{mol:T2}
			\begin{gathered}
				T_{2}: <S_{0}, 1>\mapsto <S_{2},\beta,\Rightarrow> \\
				\left[ \overrightarrow{BsrDI}|4_{F}|6_{\beta}|
				4_{F}|6|\overleftarrow{BserI}|\overrightarrow{FokI}|3|4_{F}|12|
				\overleftarrow{BpmI}|\overrightarrow{BpmI}|8|6_{1}|6|\overleftarrow{BbvI}
        				\right]
			\end{gathered}
		\end{equation}

		\begin{equation}
			\label{mol:T3}
			\begin{gathered}
				T_{3}: <S_{0}, \beta>\mapsto <HALT> \\
				\left[ \overrightarrow{BsrDI}|4_{F}|~H~A~L~T~|6_{\beta}|6|\overleftarrow{BbvI}
				\right]
			\end{gathered}
		\end{equation}

		\begin{equation}
			\label{mol:T4}
			\begin{gathered}
				T_{4}: <S_{1}, 0>\mapsto <S_{0},1,\Rightarrow> \\
				\left[ \overrightarrow{BsrDI}|4_{F}|6_{1}|
				4_{F}|6|\overleftarrow{BserI}|\overrightarrow{FokI}|5|4_{F}|12|
				\overleftarrow{BpmI}|\overrightarrow{BpmI}|8|6_{0}|7|\overleftarrow{BbvI}
				\right]
			\end{gathered}
		\end{equation}

		\begin{equation}
			\label{mol:T5}
			\begin{gathered}
				T_{5}:<S_{1}, 1>\mapsto <S_{0},1,\Rightarrow> \\
				\left[ \overrightarrow{BsrDI}|4_{F}|6_{1}|
				4_{F}|6|\overleftarrow{BserI}|\overrightarrow{FokI}|5|4_{F}|12|
				\overleftarrow{BpmI}|\overrightarrow{BpmI}|8|6_{1}|7|\overleftarrow{BbvI}
				\right]
			\end{gathered}
		\end{equation}

		\begin{equation}
			\label{mol:T6}
			\begin{gathered}
				T_{6}:<S_{1}, \beta>\mapsto <S_{0},\varepsilon,\Rightarrow> \\
				\left[ \overrightarrow{BsrDI}|4_{F}|6_{\varepsilon}|
				4_{F}|6|\overleftarrow{BserI}|\overrightarrow{FokI}|5|4_{F}|12|
				\overleftarrow{BpmI}|\overrightarrow{BpmI}|8|6_{\beta}|7|\overleftarrow{BbvI}
				\right]
			\end{gathered}
		\end{equation}

		\begin{equation}
			\label{mol:T7}
			\begin{gathered}
				T_{7}:<S_{2}, 0>\mapsto <S_{0},1,\Rightarrow> \\
				\left[ \overrightarrow{BsrDI}|4_{F}|6_{1}|
				4_{F}|6|\overleftarrow{BserI}|\overrightarrow{FokI}|5|4_{F}|12|
				\overleftarrow{BpmI}|\overrightarrow{BpmI}|8|6_{0}|8|\overleftarrow{BbvI}
				\right]
			\end{gathered}
		\end{equation}

		\begin{equation}
			\label{mol:T8}
			\begin{gathered}
				T_{8}:<S_{2}, 1>\mapsto <S_{0},0,\Rightarrow> \\
				\left[ \overrightarrow{BsrDI}|4_{F}|6_{1}|
				4_{F}|6|\overleftarrow{BserI}|\overrightarrow{FokI}|5|4_{F}|12|
				\overleftarrow{BpmI}|\overrightarrow{BpmI}|8|6_{0}|8|\overleftarrow{BbvI}
				\right]
			\end{gathered}
		\end{equation}

		\begin{equation}
			\label{mol:T9}
			\begin{gathered}
				T_{9}: <S_{2}, \beta>\mapsto <S_{0},\varepsilon,\Rightarrow> \\
				\left[ \overrightarrow{BsrDI}|4_{F}|6_{\varepsilon}|
				4_{F}|6|\overleftarrow{BserI}|\overrightarrow{FokI}|5|4_{F}|12|
				\overleftarrow{BpmI}|\overrightarrow{BpmI}|8|6_{\beta}|8|\overleftarrow{BbvI}
				\right]
			\end{gathered}
		\end{equation}

		Where the string
		
		\begin{equation}
			\left[~H~A~L~T~\right]
		\end{equation}
		
	 	denotes the double-stranded DNA molecule that acts as detection molecule for the TM, and indicates that the computation is over.
		The choice of base pairs that comprise this molecule is arbitrary as long as it does not interfere with the action of the
		restriction enzymes. As done previously, we group these molecules in a mathematical set that we name $\Gamma$.
				
		The computation begins when the circular double-stranded DNA molecule, in which we encode the input string, is mixed with all
		the transition molecules and the restriction enzymes. The \emph{head} of the machine is represented by the recognition sites of
		restriction enzymes FokI and BserI.
		
		The computation takes place immediately when the restriction enzymes react
		to the contents of the circular double-stranded DNA molecule, and the transition molecules bind to it via their sticky-ends.
		At the end of the computation, the contents of the tape can be read on the DNA strand, and this corresponds
		to the output of the TM.  
				
		After having described the contents of the sets $\Gamma$ (transition molecules), $\Phi$ (restriction enzymes) and $\Sigma$ (alphabet molecules)
		we are able to announce our main theorem.
		
		\begin{theorem}
			$<\Gamma, \Phi, \Sigma>$ is a molecular Turing machine that computes a logical NAND function.
		\end{theorem}
		
		The proof of this theorem can be found in the Appendix section of this paper, which basically proves that our design of transition molecules, alphabet and
		restriction enzymes achieves to mimic the action of a logical NAND gate on a molecular TM. Thus, showing that it is possible to build a molecular
		version of a logic unit, which together with an arithmetic unit, is an essential part of a central processing unit of any computing device.
		
\begin{figure}[h!]
			\centering
			\includegraphics[width=0.6\textwidth, height=20cm]{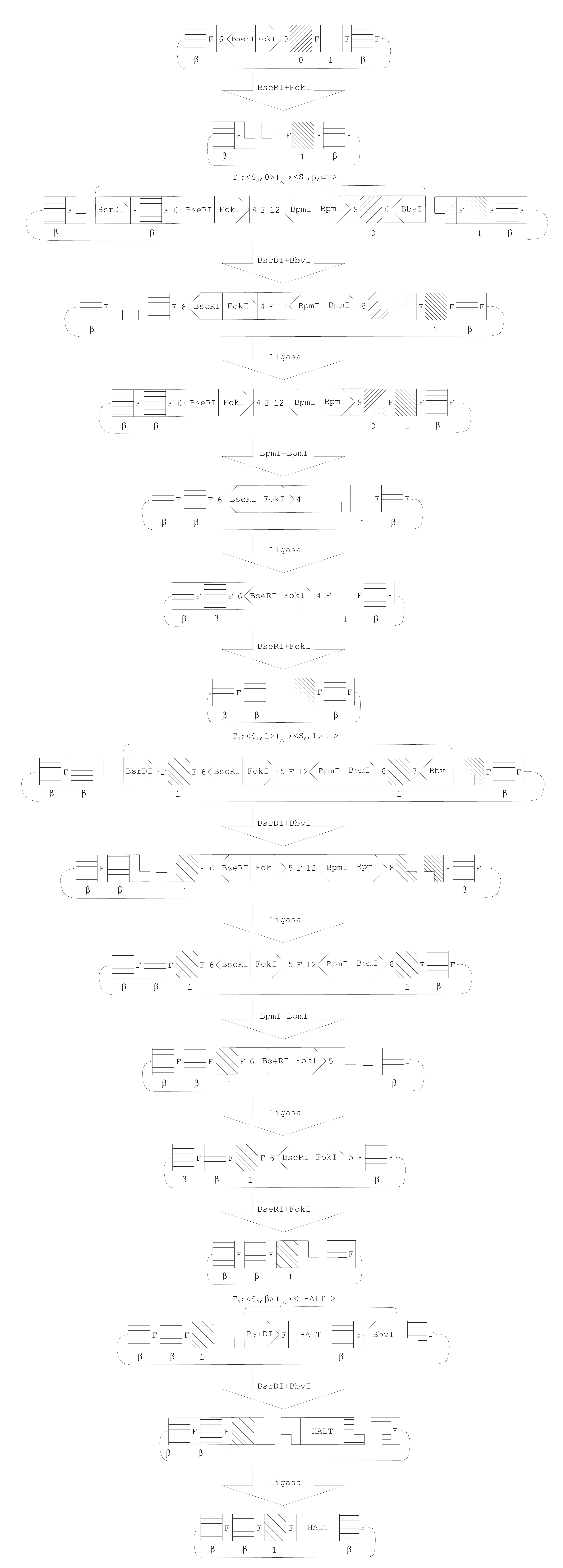}
			\caption{Molecular TM working on input \texttt{01}}
   	 		\label{fig:example}			
		\end{figure}

		Figure (\ref{fig:example}) in page \pageref{fig:example} shows an schematic example of the TM operation with input \texttt{01}, where we show how enzymes 
		cleave the molecule, and how the transition molecules bind to it mimicking the action of the head reading/writing on the tape and then moving along it.
		For input \texttt{01} we expect the machine to return \texttt{1} as output (see table (\ref{table1}) above), which is the case for our design.
		
		The action of our TM can be summarised in the following steps, naturally we are assuming ideal conditions, which means that our molecules are not affected
		by other substances in the medium and computation is carried out flawlessly:
		
		\begin{enumerate}
			\item Enzymes FokI and BserI recognise their binding sites and cleave the molecule accordingly. Because the molecule is a circular DNA strand, no nucleotide
			is lost after cleavage.
			\item Transition molecules enter the medium after being cleaved by enzyme BbvI.
			\item In the medium, enzyme BsrDI acts upon the transition molecules, which makes them susceptible to bind to the main strand.
			\item The enzyme ligase aids the coupling of transition molecules and the main strand in virtue of their complementary sticky-ends.
			\item To delete the character that has been read from the tape, enzyme BpmI acts upon the DNA strand, which cleaves that particular section of the string.
			\item Once more, enzyme ligase bind together the complementary sticky-ends, which results in a complete circular double-stranded DNA molecule.
			\item The first step is repeated until no recognition site is found within the main molecule, which implies that the string representing the \emph{HALT} state
			is contained in the molecule.
		\end{enumerate} 
	
	\section{Discussion}
		\label{sec:CONC}
		An important factor that must be taken in consideration when implementing this kind of molecular devices in the lab is the sequence of steps in which computation takes 
		place. In our mathematical abstraction, these steps occur one at a time and without errors on a unique DNA molecule. 
		In reality, we would have thousands of these molecules, plus restriction enzymes and other molecules required for the computation (e.g. ligase and ATP) in a single
		test tube, all of them influencing each other in one way or the other. Therefore, we would expect that only a percentage of those strands will perform the computation
		as planned. 
		
		Should this fact discourage us? We should keep in mind that the molecular computing paradigm is still in its infancy, and a lot of work must be done both in the theoretical
		and in the experimental fronts.
		Research efforts must be directed towards developing mechanisms for error detection and fault tolerance. In \cite{rothemund1996dna}, the author presents a series
		of strategies to deal with undesired effects during molecular runtime. Examples of these adverse effects are: faulty or incorrect bonds, and faulty or incorrect cuts.
		Strategies to deal with them include the use of exonucleases and endonucleases with no specific restriction site to detect and correct errors during run-time 
		\cite{rothemund1996dna}.
		
		Genetic material is a promising computing medium due to its relative ease of parallelisation and its natural tendency to bind to complementary strings.
		This features offer the possibility of solving hard computing problems with less effort and cost than it would require to conventional computation.
		The cost of building and maintaing a computer cluster is very high in terms of space, energy and other human and technical resources. 
		On the other hand, molecular computing offers the possibility of having thousands of molecules computing in a space as small as a teaspoon, without any negative
		environmental impact.
		
		Nevertheless, it is very likely that up to now this computer paradigm might not perform any better than our modern computers. 
		However, we should take a few steps back and think about
		the nature of this new computing paradigm. Genetic material as a computing device opens a new perspective on the kind of problems that we could face with it.
		DNA is the native language of the cell, therefore using the former as software and the latter as hardware gives us a new opportunity to deal with medical conditions
		that remain intractable to date. One of the first steps towards this direction is the construction of an arithmetic and logic unit made entirely from DNA molecules.
		Here we offered an insight for such a purpose.

	\bibliographystyle{splncs}
	\bibliography{biblio}	

	\section*{Appendix}
	
		\begin{Theo}
			$<\Gamma, \Phi, \Sigma>$ is a molecular Turing machine that computes a logical NAND function.
		\end{Theo}
		\begin{proof}
			We proceed in the following way. Let us consider a string as long as the proof requires, that is, a circular double-stranded DNA molecule
			with undefined length. Thus, we have the following initial configuration:
			\begin{equation}
				\label{teo:conf_ini}
				\left(
				6_{\beta}|4_{F}|6|\BserI|\FokI|9|6_{\lambda_{1}}|4_{F}|6_{\lambda_{2}}|4_{F}|\cdots|6_{\lambda_{i}}|4_{F}|\cdots
				\right)
			\end{equation}			
			
			At this point, the head of the machine is \emph{placed over} the character $\lambda_{1}$. The next step occurs when enzymes FokI and
			BserI recognise their sites and cleave the molecule. This action leaves the molecule in the following configuration:
			\begin{equation}
				\label{teo:paso2}
				\left(
				6_{\beta}|\frac{\scriptscriptstyle F_{1}F_{2}}{}|~|\frac{\scriptscriptstyle \lambda_{11}\lambda_{12}\lambda_{13}
				\lambda_{14}\lambda_{15}\lambda_{16}}{\scriptscriptstyle
				~~~~~~~~~~~~~
				\widetilde{\lambda_{15}}\widetilde{\lambda_{16}}}|4_{F}|6_{\lambda_{2}}|4_{F}|\cdots
				\right)
			\end{equation}			
			
			The sticky-end  $\lambda_{11}\lambda_{12}\lambda_{13}\lambda_{14}$ indicates that the machine is currently in state $S_{0}$
			with input $\lambda_{1}$. Because the molecule is circular, that is, it is linked by its extremes, the computation will proceed and
			a molecule has a chance to fill the gap caused by the action of the enzymes, provided it has a complementary sticky-end.
			Let us analyse the different cases for the value of character $\lambda_{1}$.
			
			\begin{description}
				\item[CASE 1: $<S_{0}, \lambda_{1}=\beta>$] Then, string (\ref{teo:paso2}) is actually the string:
					\begin{equation}
						\label{teo:paso3}
						\left(
						6_{\beta}|\frac{\scriptscriptstyle F_{1}F_{2}}{}|~|\frac{\scriptscriptstyle \beta_{1}\beta_{2}\beta_{3}
						\beta_{4}\beta_{5}\beta_{6}}{\scriptscriptstyle ~~~~~~~~~~
						\widetilde{\beta_{5}}\widetilde{\beta_{6}}}|4_{F}|6_{\lambda_{2}}|4_{F}|\cdots
						\right)
					\end{equation}
					And thus, we expect transition $T_{3}$  (\ref{mol:T3}) to take place. 
					The molecule representing such transition gets close to the empty space in string (\ref{teo:paso3}) as shown below:
					\begin{equation}
						\begin{gathered}
							\left(
							6_{\beta}|\frac{\scriptscriptstyle
							F_{1}F_{2}}{}|~~~~~~~~~~~~~~~~~~~~~~~~~~~~~~~~~~~~~~|\frac{\scriptscriptstyle \beta_{1}\beta_{2}\beta_{3}
							\beta_{4}\beta_{5}\beta_{6}}{\scriptscriptstyle ~~~~~~~~~~
							\widetilde{\beta_{5}}\widetilde{\beta_{6}}}|4_{F}|\cdots
							\right)
							\\
							|\BsrDI|4_{F}|~H~A~L~T~|6_{\beta}|6|\BbvI|~~~~~~~~~~~~~~~
						\end{gathered}
					\end{equation}	
					
					However, this molecule cannot incorporate to the main strand yet, as it does not exhibit any sticky-end.
					This happens after enzymes BsrDI and BbvI act on it, which results in the following string:
					\begin{equation}
						\label{teo:paso5}
						\begin{gathered}
							\left(
							6_{\beta}|\frac{\scriptscriptstyle
							F_{1}F_{2}}{}|~~~~~~~~~~~~~~~~~~~~~~~~~~~~~~~~~~~~~~|\frac{\scriptscriptstyle \beta_{1}\beta_{2}\beta_{3}
							\beta_{4}\beta_{5}\beta_{6}}{\scriptscriptstyle ~~~~~~~~~~
							\widetilde{\beta_{5}}\widetilde{\beta_{6}}}|4_{F}|\cdots\right)
							\\
							|\frac{\scriptscriptstyle ~~~~~F_{3}F_{4}}{\scriptscriptstyle \widetilde{F_{1}}\widetilde{F_{2}}
							\widetilde{F_{3}}\widetilde{F_{4}}}|~H~A~L~T~|\frac{}{\scriptscriptstyle \widetilde{\beta_{1}}\widetilde{\beta_{2}}
							\widetilde{\beta_{3}}\widetilde{\beta_{4}}}|~~~~~~~~~~~~~~~
						\end{gathered}
					\end{equation}	
					
					As the sticky-ends in both molecules are complementary, they are susceptible of binding together, which results in the
					following molecule:
					\begin{equation}
						\label{teo:paso6}
						\left(
						6_{\beta}|4_{F}|~H~A~L~T~|6_{\beta}|4_{F}|\cdots
						\right)
					\end{equation}
					
					Resulting in a complete molecule with no recognition sites on which an enzyme could act. Thus, at this stage the computation
					ends as expected.		
				\item[CASE 2: $<S_{0}, \lambda_{1}=0>$] Thus, string (\ref{teo:paso2}) is actually the string:
					\begin{equation}
						\label{teo:paso7}
						\left(
						6_{\beta}|\frac{\scriptscriptstyle F_{1}F_{2}}{}|~|\frac{\scriptscriptstyle A_{1}A_{2}A_{3}
						A_{4}A_{5}A_{6}}{\scriptscriptstyle ~~~~~~~~~~
						\widetilde{A_{5}}\widetilde{A_{6}}}|4_{F}|6_{\lambda_{2}}|4_{F}|\cdots
						\right)
					\end{equation}
					In this case, transition $T_{1}$ (\ref{mol:T1}) should take place, which at the moment of attempting to bind to the main
					molecule results in:
					\begin{equation}
						\begin{gathered}
							\left( 6_{\beta}|\frac{\scriptscriptstyle
							F_{1}F_{2}}{}|~~~~~~~~~~~~~~~~~~~~~~~~~~~~~~~~~~~~~~~~~~~~~~~~~~~~~~~~~~~~~~~~~~~~~~
							|\frac{\scriptscriptstyle
							A_{1}A_{2}A_{3}A_{4}A_{5}A_{6}}{\scriptscriptstyle
							~~~~~~~~~~~~\widetilde{A_{5}}\widetilde{A_{6}}}|
							\cdots
							\right)
							\\
							|\overrightarrow{\scriptscriptstyle BsrDI}|4_{F}|6_{\beta}|
							4_{F}|6|\overleftarrow{\scriptscriptstyle BserI}|\overrightarrow{\scriptscriptstyle FokI}|4|4_{F}|12|
							\overleftarrow{\scriptscriptstyle BpmI}|\overrightarrow{\scriptscriptstyle BpmI}|8|6_{0}|6|
							\overleftarrow{\scriptscriptstyle BbvI}|~~~~~~~~~~~~~~
						\end{gathered}
					\end{equation}
					
					After enzymes BsrDI and BbvI act upon transition molecule $T_{1}$ they yield sticky-ends in both extremes which
					are susceptible of binding to the main molecule, which results in:
					\begin{equation}
						\begin{gathered}
							\left( 6_{\beta}|\frac{\scriptscriptstyle F_{1}F_{2}}{}|~~~~~~~~~~~~~~~~~~~~~~~~~~~~~~~~~~~~~~~~~~~~~~~~~~~~~~~~~~~~~~~~~~~~~~
							|\frac{\scriptscriptstyle
							A_{1}A_{2}A_{3}A_{4}A_{5}A_{6}}{\scriptscriptstyle
							~~~~~~~~~~~~\widetilde{A_{5}}\widetilde{A_{6}}}|
							\cdots
							\right)
							\\
							|\frac{\scriptscriptstyle ~~~~~F_{3}F_{4}}{\scriptscriptstyle
							\widetilde{F_{1}}\widetilde{F_{2}}\widetilde{F_{3}}\widetilde{F_{4}}}|6_{\beta}|
							4_{F}|6|\overleftarrow{\scriptscriptstyle BserI}|\overrightarrow{\scriptscriptstyle FokI}|4|4_{F}|12|
							\overleftarrow{\scriptscriptstyle BpmI}|\overrightarrow{\scriptscriptstyle BpmI}|8|\frac{}{\scriptscriptstyle
							\widetilde{A_{1}}\widetilde{A_{2}}\widetilde{A_{3}}\widetilde{A_{4}}}|~~~~~~~~~~~~~~
						\end{gathered}
					\end{equation}
					
					And then,
					\begin{equation}
						\label{teo:paso10}
						\left( 6_{\beta}|4_{F}|6_{\beta}|
						4_{F}|6|\overleftarrow{\scriptscriptstyle BserI}|\overrightarrow{\scriptscriptstyle FokI}|4|4_{F}|12|
						\overleftarrow{\scriptscriptstyle BpmI}|\overrightarrow{\scriptscriptstyle
						BpmI}|8|6_{0}|4_{F}|6_{\lambda_{2}}|4_{F}|
						\cdots
						\right)
					\end{equation}
					
					In the next step, the enzyme BpmI recognises its site and binds to the molecule cleaving it, which results in:
					\begin{equation}
						\left(
						6_{\beta}|4_{F}|6_{\beta}|4_{F}|6|\BserI|\FokI|4|\frac{\scriptscriptstyle
						F_{1}F_{2}}{}|~|\frac{\scriptscriptstyle ~~~~~F_{3}F_{4}}{\scriptscriptstyle \widetilde{F_{1}}\widetilde{F_{2}}
						\widetilde{F_{3}}\widetilde{F_{4}}}|6_{\lambda_{2}}|4_{F}|
						\cdots
						\right)
					\end{equation}
					
					where the sticky-ends are susceptible of binding, which yields the following configuration:
					\begin{equation}
						\left(
						6_{\beta}|4_{F}|6_{\beta}|4_{F}|6|\BserI|\FokI|4|4_{F}|6_{\lambda_{2}}|4_{F}|
						\cdots
						\right)
					\end{equation}
					
					which we consider as a new initial configuration for the machine, as we did for the configuration (\ref{teo:conf_ini}).
					At this point, the head is \emph{placed over} the character $\lambda_{2}$. We will continue exploring this branch
					of the computation later in this proof (see below). We continue developing the different cases of $\lambda_{1}$.
				\item[CASE 3: $<S_{0}, \lambda_{1}=1>$] So, the string (\ref{teo:paso2}) is actually the following string:
					\begin{equation}
						\label{teo:paso13}
						\left(
						6_{\beta}|\frac{\scriptscriptstyle F_{1}F_{2}}{}|~|\frac{\scriptscriptstyle B_{1}B_{2}B_{3}
						B_{4}B_{5}B_{6}}{\scriptscriptstyle ~~~~~~~~~~
						\widetilde{B_{5}}\widetilde{B_{6}}}|4_{F}|6_{\lambda_{2}}|4_{F}|\cdots
						\right)
					\end{equation}
					
					At this moment, transition $T_{2}$ should take place. Such a molecule (\ref{mol:T2}) attempts to bind to the main molecule
					which results in the following:
					\begin{equation}
						\begin{gathered}
							\left(
							6_{\beta}|\frac{\scriptscriptstyle
							F_{1}F_{2}}{}|~~~~~~~~~~~~~~~~~~~~~~~~~~~~~~~~~~~~~~~~~~~~~~~~~~~~~~~~~~~~~~~~~~~~~~
							|\frac{\scriptscriptstyle B_{1}B_{2}B_{3}
							B_{4}B_{5}B_{6}}{\scriptscriptstyle ~~~~~~~~~~~
							\widetilde{B_{5}}\widetilde{B_{6}}}|\cdots
							\right)
							\\
							|\BsrDI|4_{F}|6_{\beta}|4_{F}|6|\BserI|\FokI|3|4_{F}|12|\BpmII|\BpmID|8|6_{1}|6|\BbvI|~~~~~~~~~~~~~
						\end{gathered}
					\end{equation}
					
					After enzymes BsrDI and BbvI act upon this transition molecule, we obtain the following pair of strings:
					\begin{equation}
						\begin{gathered}
							\left(
							6_{\beta}|\frac{\scriptscriptstyle
							F_{1}F_{2}}{}|~~~~~~~~~~~~~~~~~~~~~~~~~~~~~~~~~~~~~~~~~~~~~~~~~~~~~~~~~~~~~~~~~~~~~~
							|\frac{\scriptscriptstyle B_{1}B_{2}B_{3}
							B_{4}B_{5}B_{6}}{\scriptscriptstyle ~~~~~~~~~~~
							\widetilde{B_{5}}\widetilde{B_{6}}}|\cdots
							\right)
							\\
							|\frac{\scriptscriptstyle ~~~~~F_{3}F_{4}}{\scriptscriptstyle
							\widetilde{F_{1}}\widetilde{F_{2}}\widetilde{F_{3}}\widetilde{F_{4}}}|6_{\beta}|4_{F}|6|\BserI|\FokI|3|4_{F}|12|
							\BpmII|\BpmID|8|\frac{}{\scriptscriptstyle \widetilde{B_{1}}\widetilde{B_{2}}\widetilde{B_{3}}\widetilde{B_{4}}}|
							~~~~~~~~~~~~~
						\end{gathered}
					\end{equation}
					
					The sticky-ends of the last molecule are susceptible of binding together, which results in the following molecule:
					\begin{equation}
						\label{teo:paso16}
						\left(
						6_{\beta}|4_{F}|6_{\beta}|4_{F}|6|\BserI|\FokI|3|4_{F}|12|
						\BpmII|\BpmID|8|6_{1}|4_{F}|6_{\lambda_{2}}|4_{F}|\cdots
						\right)
					\end{equation}
					
					Next, enzyme BpmI recognises its binding site and cleaves the molecule resulting in:
					\begin{equation}
						\left(
						6_{\beta}|4_{F}|6_{\beta}|4_{F}|6|\BserI|\FokI|3|\frac{\scriptscriptstyle
						F_{1}F_{2}}{}|~|\frac{\scriptscriptstyle ~~~~~F_{3}F_{4}}{\scriptscriptstyle
						\widetilde{F_{1}}\widetilde{F_{2}}\widetilde{F_{3}}\widetilde{F_{4}}}|
						6_{\lambda_{2}}|4_{F}|\cdots
						\right)
					\end{equation}
					
					The sticky-ends of the last molecule are susceptible of binding together, which yields the following configuration:
					\begin{equation}
						\left(
						6_{\beta}|4_{F}|6_{\beta}|4_{F}|6|\BserI|\FokI|3|4_{F}|
						6_{\lambda_{2}}|4_{F}|\cdots
						\right)
					\end{equation}
					
					which analogously to \textbf{CASE 2} is a new configuration for the machine, whose development will be studied
					below. At this moment, the head is reading character $\lambda_{2}$ after writing character $\beta$ as expected.
			\end{description}			
			
			With our previous exposition, we have exhausted all the possible cases for character $\lambda_{1}$, we proceed now to
			explore the branches of the computation that resulted from developing the first character.
			Given that \textbf{CASE 1} represents a state in which the machine has stopped, we analyse the other two remaining case
			and branch accordingly.
			Now, we explore the \textbf{CASE 2} branch, that is, when $\lambda_{1} = 0$, we have as initial configuration the strand:
			\begin{equation}
				\left(
				6_{\beta}|4_{F}|6_{\beta}|4_{F}|6|\BserI|\FokI|4|4_{F}|6_{\lambda_{2}}|4_{F}|
				\cdots
				\right)
			\end{equation}
			
			The head is currently over the character $\lambda_{2}$. When enzymes FokI and BserI act upon this string, they yield the
			following string:
			\begin{equation}
				\label{teo:paso20}
				\left(
				6_{\beta}|4_{F}|6_{\beta}|\frac{\scriptscriptstyle F_{1}F_{2}}{}|~|\frac{\scriptscriptstyle \lambda_{22}\lambda_{23}
				\lambda_{24}\lambda_{25}\lambda_{26}}{\scriptscriptstyle
				~~~~~~~~~~~~~~\widetilde{\lambda_{26}}}|4_{F}|6_{\lambda_{3}}|4_{F}|\cdots
				\right)
			\end{equation}
			
			where the sticky-end $\lambda_{22}\lambda_{23}\lambda_{24}\lambda_{25}$ reports that the machine is currently in state
			$S_{1}$ with input $\lambda_{2}$, which was expected from the transition rules after reading character $0$ on the tape.
			Let us explore the different cases for $\lambda_{2}$.
			
			\begin{description}
				\item[CASE 2.1: $<S_{1}, \lambda_{2}=\beta>$] So, the string (\ref{teo:paso20}) is actually the string:
					\begin{equation}
						\label{teo:paso21}
						\left(
						6_{\beta}|4_{F}|6_{\beta}|\frac{\scriptscriptstyle F_{1}F_{2}}{}|~|\frac{\scriptscriptstyle \beta_{2}\beta_{3}
						\beta_{4}\beta_{5}\beta_{6}}{\scriptscriptstyle
						~~~~~~~~~\widetilde{\beta_{6}}}|4_{F}|6_{\lambda_{3}}|4_{F}|\cdots
						\right)
					\end{equation}
					According to our transition rules, when the machine reads character $\beta$ in state $S_{1}$, transition $T_{6}$ takes place
					that is, molecule (\ref{mol:T6}) attempts to bind to the main molecule, which looks like:
					\begin{equation}
						\begin{gathered}
							\left(
							6_{\beta}|4_{F}|6_{\beta}|\frac{\scriptscriptstyle F_{1}F_{2}}{}|
							~~~~~~~~~~~~~~~~~~~~~~~~~~~~~~~~~~~~~~~~~~~~~~~~~~~~~~~~~~~~~~~~~~~~~~
							|\frac{\scriptscriptstyle \beta_{2}\beta_{3}
							\beta_{4}\beta_{5}\beta_{6}}{\scriptscriptstyle
							~~~~~~~~~\widetilde{\beta_{6}}}|\cdots
							\right)
							\\
							|\BsrDI|4_{F}|6_{\varepsilon}|4_{F}|6|
							\BserI|\FokI|5|4_{F}|12|\BpmII|\BpmID|8|6_{\beta}|7|\BbvI|
						\end{gathered}
					\end{equation}
					Once the enzymes BsrDI and BbvI have cleaved the transition molecule, the string becomes:
					\begin{equation}
						\begin{gathered}
							\left(
							6_{\beta}|4_{F}|6_{\beta}|\frac{\scriptscriptstyle F_{1}F_{2}}{}|
							~~~~~~~~~~~~~~~~~~~~~~~~~~~~~~~~~~~~~~~~~~~~~~~~~~~~~~~~~~~~~~~~~~~~~~
							|\frac{\scriptscriptstyle \beta_{2}\beta_{3}\beta_{4}\beta_{5}\beta_{6}}{\scriptscriptstyle
							~~~~~~~~~\widetilde{\beta_{6}}}|\cdots
							\right)
							\\
							|\frac{\scriptscriptstyle ~~~~~F_{3}F_{4}}{\scriptscriptstyle
							\widetilde{F_{1}}\widetilde{F_{2}}\widetilde{F_{3}}\widetilde{F_{4}}}|6_{\varepsilon}|4_{F}|6|
							\BserI|\FokI|5|4_{F}|12|\BpmII|\BpmID|8|\frac{\scriptscriptstyle \beta_{1}~~~~~~~~~}{\scriptscriptstyle
							\widetilde{\beta_{1}}\widetilde{\beta_{2}}\widetilde{\beta_{3}}\widetilde{\beta_{4}}\widetilde{\beta_{5}}}|
						\end{gathered}
					\end{equation}
				
					Here, the sticky-ends are susceptible of binding together. When this happens, the string becomes:
					\begin{equation}
						\left(
						6_{\beta}|4_{F}|6_{\beta}|4_{F}|6_{\varepsilon}|4_{F}|6|
						\BserI|\FokI|5|4_{F}|12|\BpmII|\BpmID|8|6_{\beta}|4_{F}|6_{\lambda_{3}}|4_{F}|\cdots
						\right)
					\end{equation}
					
					Then, enzyme BpmI cleaves the molecule, which results in the following:
					\begin{equation}
						\left(
						6_{\beta}|4_{F}|6_{\beta}|4_{F}|6_{\varepsilon}|4_{F}|6|
						\BserI|\FokI|5|\frac{\scriptscriptstyle F_{1}F_{2}}{}|~|\frac{\scriptscriptstyle ~~~~~F_{3}F_{4}}{\scriptscriptstyle
						\widetilde{F_{1}}\widetilde{F_{2}}\widetilde{F_{3}}\widetilde{F_{4}}}|6_{\lambda_{3}}|4_{F}|\cdots
						\right)
					\end{equation}
					
					We see sticky-ends that are susceptible of binding together. When this happens the resulting molecule is:
					\begin{equation}
						\label{teo:paso26}
						\left(
						6_{\beta}|4_{F}|6_{\beta}|4_{F}|6_{\varepsilon}|4_{F}|6|
						\BserI|\FokI|5|4_{F}|6_{\lambda_{3}}|4_{F}|\cdots
						\right)
					\end{equation}
					
					Up to this point, we know that the head is placed over the character $\lambda_{3}$ and it has written the symbol
					$\varepsilon$ on the tape. This string can be considered as a new initial configuration for the machine, but before
					we continue exploring its branch, we take a look at the remaining cases.
				\item[CASE 2.2: $<S_{1}, \lambda_{2}=0>$] So, the string (\ref{teo:paso20}) is actually the string:
					\begin{equation}
						\left(
						6_{\beta}|4_{F}|6_{\beta}|\frac{\scriptscriptstyle F_{1}F_{2}}{}|~|\frac{\scriptscriptstyle A_{2}A_{3}
						A_{4}A_{5}A_{6}}{\scriptscriptstyle
						~~~~~~~~~~~\widetilde{A_{6}}}|4_{F}|6_{\lambda_{3}}|4_{F}|\cdots
						\right)
					\end{equation}
					Given the current state of the machine and the character being read, the transition $T_{4}$ must take place, and thus
					molecule (\ref{mol:T4}) attempts to add to the main molecule:
					\begin{equation}
						\begin{gathered}
							\left(
							6_{\beta}|4_{F}|6_{\beta}|\frac{\scriptscriptstyle F_{1}F_{2}}{}|
							~~~~~~~~~~~~~~~~~~~~~~~~~~~~~~~~~~~~~~~~~~~~~~~~~~~~~~~~~~~~~~~~~~~~~~
							|\frac{\scriptscriptstyle A_{2}A_{3}
							A_{4}A_{5}A_{6}}{\scriptscriptstyle
							~~~~~~~~~~~\widetilde{A_{6}}}|\cdots
							\right)
							\\
							|\BsrDI|4_{F}|6_{1}|4_{F}|6|
							\BserI|\FokI|5|4_{F}|12|\BpmII|\BpmID|8|6_{0}|7|\BbvI|~~
						\end{gathered}
					\end{equation}
					
					When enzymes BsrDI and BbvI act upon the transition molecule they produce the following string:
					\begin{equation}
						\begin{gathered}
							\left(
							6_{\beta}|4_{F}|6_{\beta}|\frac{\scriptscriptstyle F_{1}F_{2}}{}|
							~~~~~~~~~~~~~~~~~~~~~~~~~~~~~~~~~~~~~~~~~~~~~~~~~~~~~~~~~~~~~~~~~~~~~~
							|\frac{\scriptscriptstyle A_{2}A_{3}
							A_{4}A_{5}A_{6}}{\scriptscriptstyle
							~~~~~~~~~~~\widetilde{A_{6}}}|\cdots
							\right)
							\\
							|\frac{\scriptscriptstyle ~~~~~F_{3}F_{4}}{\scriptscriptstyle
							\widetilde{F_{1}}\widetilde{F_{2}}\widetilde{F_{3}}\widetilde{F_{4}}}|6_{1}|4_{F}|6|
							\BserI|\FokI|5|4_{F}|12|\BpmII|\BpmID|8|\frac{\scriptscriptstyle A_{1}~~~~~~~~~~~}{\scriptscriptstyle
							\widetilde{A_{1}}\widetilde{A_{2}}\widetilde{A_{3}}\widetilde{A_{4}}\widetilde{A_{5}}}|~~
						\end{gathered}
					\end{equation}
					
					which has sticky-ends susceptible of binding together. When they do so, the resulting molecule is:
					\begin{equation}
						\left(
						6_{\beta}|4_{F}|6_{\beta}|4_{F}|6_{1}|4_{F}|6|
						\BserI|\FokI|5|4_{F}|12|\BpmII|\BpmID|8|6_{0}|4_{F}|6_{\lambda_{3}}|4_{F}|\cdots
						\right)
					\end{equation}
					
					Next, the enzyme BpmI cleaves the molecule, which results in:
					\begin{equation}
						\left(
						6_{\beta}|4_{F}|6_{\beta}|4_{F}|6_{1}|4_{F}|6|
						\BserI|\FokI|5|\frac{\scriptscriptstyle F_{1}F_{2}}{}|~|\frac{\scriptscriptstyle ~~~~~F_{3}F_{4}}{\scriptscriptstyle
						\widetilde{F_{1}}\widetilde{F_{2}}\widetilde{F_{3}}\widetilde{F_{4}}}|6_{\lambda_{3}}|4_{F}|\cdots
						\right)
					\end{equation}
					
					Again, the molecule has sticky-ends susceptible of binding together. Thus, the molecule becomes:
					\begin{equation}
						\label{teo:paso32}
						\left(
						6_{\beta}|4_{F}|6_{\beta}|4_{F}|6_{1}|4_{F}|6|
						\BserI|\FokI|5|4_{F}|6_{\lambda_{3}}|4_{F}|\cdots
						\right)
					\end{equation}
					At this point, the head is placed over the character $\lambda_{3}$ after writing character $1$ on the tape., which is
					expected after reading characters $00$ on the tape. The new state is $S_{0}$ and below we describe what happens
					if we further elaborate this case.
				\item[CASE 2.3: $<S_{1}, \lambda_{2}=1>$] Then, string (\ref{teo:paso20}) is actually the string:
					\begin{equation}
						\label{teo:paso33}
						\left(
						6_{\beta}|4_{F}|6_{\beta}|\frac{\scriptscriptstyle F_{1}F_{2}}{}|~|\frac{\scriptscriptstyle B_{2}B_{3}
						B_{4}B_{5}B_{6}}{\scriptscriptstyle
						~~~~~~~~~~~\widetilde{B_{6}}}|4_{F}|6_{\lambda_{3}}|4_{F}|\cdots
						\right)
					\end{equation}
					In this situation, the machine is in state $S_{1}$ and input $\lambda_{2} = 1$, thus transition $T_{5}$ should take
					place. When the molecule representing this transition (\ref{mol:T5}) attempts to bind to the main molecule, we obtain:
					\begin{equation}
						\begin{gathered}
							\left(
							6_{\beta}|4_{F}|6_{\beta}|\frac{\scriptscriptstyle F_{1}F_{2}}{}|
							~~~~~~~~~~~~~~~~~~~~~~~~~~~~~~~~~~~~~~~~~~~~~~~~~~~~~~~~~~~~~~~~~~~~~~
							|\frac{\scriptscriptstyle B_{2}B_{3}
							B_{4}B_{5}B_{6}}{\scriptscriptstyle
							~~~~~~~~~~~\widetilde{B_{6}}}|\cdots
							\right)
							\\
							|\BsrDI|4_{F}|6_{1}|4_{F}|6|
							\BserI|\FokI|5|4_{F}|12|\BpmII|\BpmID|8|6_{1}|7|\BbvI|~~
						\end{gathered}
					\end{equation}
					
					Here, enzymes BsrDI and BbvI recognise their sites and cleave the molecule, which results in:
					\begin{equation}
						\begin{gathered}
							\left(
							6_{\beta}|4_{F}|6_{\beta}|\frac{\scriptscriptstyle F_{1}F_{2}}{}|
							~~~~~~~~~~~~~~~~~~~~~~~~~~~~~~~~~~~~~~~~~~~~~~~~~~~~~~~~~~~~~~~~~~~~~~
							|\frac{\scriptscriptstyle B_{2}B_{3}
							B_{4}B_{5}B_{6}}{\scriptscriptstyle
							~~~~~~~~~~~\widetilde{B_{6}}}|\cdots
							\right)
							\\
							|\frac{\scriptscriptstyle ~~~~~F_{3}F_{4}}{\scriptscriptstyle
							\widetilde{F_{1}}\widetilde{F_{2}}\widetilde{F_{3}}\widetilde{F_{4}}}|6_{1}|4_{F}|6|
							\BserI|\FokI|5|4_{F}|12|\BpmII|\BpmID|8|\frac{\scriptscriptstyle B_{1}~~~~~~~~~~~}{\scriptscriptstyle
							\widetilde{B_{1}}\widetilde{B_{2}}\widetilde{B_{3}}\widetilde{B_{4}}\widetilde{B_{5}}}|~~
						\end{gathered}
					\end{equation}
					This allows the sticky-ends to bind together, which produces the following strand:
					\begin{equation}
						\left(
						6_{\beta}|4_{F}|6_{\beta}|4_{F}|6_{1}|4_{F}|6|
						\BserI|\FokI|5|4_{F}|12|\BpmII|\BpmID|8|6_{1}|4_{F}|6_{\lambda_{3}}|4_{F}|\cdots
						\right)
					\end{equation}
					
					Finally, the enzyme BpmI cleaves the molecules producing:
					\begin{equation}
						\left(
						6_{\beta}|4_{F}|6_{\beta}|4_{F}|6_{1}|4_{F}|6|
						\BserI|\FokI|5|\frac{\scriptscriptstyle F_{1}F_{2}}{}|~|\frac{\scriptscriptstyle ~~~~~F_{3}F_{4}}{\scriptscriptstyle
						\widetilde{F_{1}}\widetilde{F_{2}}\widetilde{F_{3}}\widetilde{F_{4}}}|6_{\lambda_{3}}|4_{F}|\cdots
						\right)
					\end{equation}
					And after the sticky-ends bind together, we obtain the molecule:
					\begin{equation}
						\label{teo:paso38}
						\left(
						6_{\beta}|4_{F}|6_{\beta}|4_{F}|6_{1}|4_{F}|6|
						\BserI|\FokI|5|4_{F}|6_{\lambda_{3}}|4_{F}|\cdots
						\right)
					\end{equation}
					
					Which can be considered a new initial configuration for the machine. The head is now placed over the character
					$\lambda_{3}$ in state $S_{0}$ after writing character $1$ on the tape, as expected after reading the string $01$.
			\end{description}
			We wonder what would happen if we further developed the recently described cases. We answer this question in the following
			observation.
			\begin{Obs}
				Let us take a look at the tapes that result from following the three last cases (\textbf{2.1},\textbf{2.2},\textbf{2.3}).
				First thing we observe is that strings (\ref{teo:paso32}) and (\ref{teo:paso38}) are the same.
				Moreover, strings (\ref{teo:paso26}), (\ref{teo:paso32}) y (\ref{teo:paso38}) correspond to state $S_{0}$ with input
				$\lambda_{3}$ differing only in the contents of its tape.
				This strings are similar to the initial configuration of the machine, string (\ref{teo:conf_ini}), and thus, we expect to
				get the same results when branching from it, no matter the current contents of the tape.
				In other words, we would expect the same molecular reactions to occur, which has been described above.
				This is valid when considering $\lambda_{1}=0$. On the other hand, if we consider $\lambda_{1}=1$ we can further
				extend the computing branch of the machine. We proceed in this direction below.
			\end{Obs}		
			
			We have the following DNA strand as initial configuration for the machine, in which the head is reading character $\lambda_{2}$.
			\begin{equation}
				\left(
				6_{\beta}|4_{F}|6_{\beta}|4_{F}|6|\BserI|\FokI|3|4_{F}|
				6_{\lambda_{2}}|4_{F}|\cdots
				\right)
			\end{equation}
			
			Enzymes FokI and BserI act upon this molecule, so that we obtain the following:
			\begin{equation}
				\label{teo:paso43}
					\left(
					6_{\beta}|4_{F}|6_{\beta}|\frac{\scriptscriptstyle F_{1}F_{2}}{}|~|\frac{\scriptscriptstyle \lambda_{23}
					\lambda_{24}\lambda_{25}\lambda_{26}}{}|4_{F}|
					6_{\lambda_{3}}|4_{F}|\cdots
					\right)
			\end{equation}
			
			Here, the sticky-end reports that the machine is currently in state $S_{2}$ with input $\lambda_{2}$.
			Once again, we explore the different cases for the value of $\lambda_{2}$.
			
			\begin{description}
				\item[CASE 3.1: $<S_{2}, \lambda_{2}=\beta>$] So, the string (\ref{teo:paso43}) is actually the string:
					\begin{equation}
						\label{teo:paso44}
						\left(
						6_{\beta}|4_{F}|6_{\beta}|\frac{\scriptscriptstyle F_{1}F_{2}}{}|~|\frac{\scriptscriptstyle \beta_{3}
						\beta_{4}\beta_{5}\beta_{6}}{}|4_{F}|
						6_{\lambda_{3}}|4_{F}|\cdots
						\right)
					\end{equation}
					Given the input $\lambda_{2}=\beta$ and state $S_{2}$ we expect transition $T_{9}$ to take place.
					This transition is given by the molecule (\ref{mol:T9}) and on attempting to bind to the main molecule
					results in:
					\begin{equation}
						\begin{gathered}
							\left(
							6_{\beta}|4_{F}|6_{\beta}|\frac{\scriptscriptstyle F_{1}F_{2}}{}|~~~~~~~~~~~~~~~~~~~~~~~~~~~~~~~~~~~~~~~~~~~~~~~~~~~~~~~~~~~~~~~~~~~~~
							|\frac{\scriptscriptstyle \beta_{3}
							\beta_{4}\beta_{5}\beta_{6}}{}|\cdots
							\right)
							\\
							|\BsrDI|4_{F}|6_{\varepsilon}|4_{F}|6|
							\BserI|\FokI|5|4_{F}|12|\BpmII|\BpmID|8|6_{\beta}|8|\BbvI|
						\end{gathered}
					\end{equation}
					After enzymes BsrDI and BbvI have recognised their sites, the molecule becomes:
					\begin{equation}
						\begin{gathered}
							\left(
							6_{\beta}|4_{F}|6_{\beta}|\frac{\scriptscriptstyle F_{1}F_{2}}{}|~~~~~~~~~~~~~~~~~~~~~~~~~~~~~~~~~~~~~~~~~~~~~~~~~~~~~~~~~~~~~~~~~~~~~~~~~
							|\frac{\scriptscriptstyle \beta_{3}
							\beta_{4}\beta_{5}\beta_{6}}{}|\cdots
							\right)
							\\
							|\frac{\scriptscriptstyle ~~~~~F_{3}F_{4}}{\scriptscriptstyle
							\widetilde{F_{1}}\widetilde{F_{2}}\widetilde{F_{3}}\widetilde{F_{4}}}|6_{\varepsilon}|4_{F}|6|
							\BserI|\FokI|5|4_{F}|12|\BpmII|\BpmID|8|\frac{\scriptscriptstyle \beta_{1}\beta_{2}~~~~~~~~~}{\scriptscriptstyle
							\widetilde{\beta_{1}}\widetilde{\beta_{2}}\widetilde{\beta_{3}}\widetilde{\beta_{4}}\widetilde{\beta_{5}}\widetilde{\beta_{6}}}
							|
						\end{gathered}
					\end{equation}
					Which binds to the main molecule producing:
					\begin{equation}
						\left(
						6_{\beta}|4_{F}|6_{\beta}|4_{F}|6_{\varepsilon}|4_{F}|6|
						\BserI|\FokI|5|4_{F}|12|\BpmII|\BpmID|8|6_{\beta}|4_{F}|6_{\lambda_{3}}|4_{F}|\cdots
						\right)
					\end{equation}
					Next, enzyme BpmI recognises its site and cleaves the molecule, thus producing:
					\begin{equation}
						\left(
						6_{\beta}|4_{F}|6_{\beta}|4_{F}|6_{\varepsilon}|4_{F}|6|
						\BserI|\FokI|5|\frac{\scriptscriptstyle F_{1}F_{2}}{}|~|\frac{\scriptscriptstyle ~~~~~F_{3}F_{4}}{\scriptscriptstyle
						\widetilde{F_{1}}\widetilde{F_{2}}\widetilde{F_{3}}\widetilde{F_{4}}}|6_{\lambda_{3}}|4_{F}|\cdots
						\right)
					\end{equation}
					When the complementary sticky-ends of this molecule bind together the molecule becomes:
					\begin{equation}
						\label{teo:paso49}
						\left(
						6_{\beta}|4_{F}|6_{\beta}|4_{F}|6_{\varepsilon}|4_{F}|6|
						\BserI|\FokI|5|4_{F}|6_{\lambda_{3}}|4_{F}|\cdots
						\right)
					\end{equation}
					
					Which is a new initial configuration for the machine, and will be analysed later in this proof, meanwhile we know that
					the head is placed over character $\lambda_{3}$ and it has just written character $\varepsilon$ on the tape as it was
					expected from input $1\beta$.
				\item[CASE 3.2: $<S_{2}, \lambda_{2}=0>$] So, string (\ref{teo:paso43}) is actually the string:
					\begin{equation}
						\left(
						6_{\beta}|4_{F}|6_{\beta}|\frac{\scriptscriptstyle F_{1}F_{2}}{}|~|\frac{\scriptscriptstyle A_{3}
						A_{4}A_{5}A_{6}}{}|4_{F}|
						6_{\lambda_{3}}|4_{F}|\cdots
						\right)
					\end{equation}
  
					For input $\lambda_{2}=0$ in state $S_{2}$ we expect transition $T_{7}$ to occur. Thus, the molecule becomes:
					\begin{equation}
						\begin{gathered}
							\left(
							6_{\beta}|4_{F}|6_{\beta}|\frac{\scriptscriptstyle F_{1}F_{2}}{}|
							~~~~~~~~~~~~~~~~~~~~~~~~~~~~~~~~~~~~~~~~~~~~~~~~~~~~~~~~~~~~~~~~~~~~~~~
							|\frac{\scriptscriptstyle A_{3}
							A_{4}A_{5}A_{6}}{}|\cdots
							\right)
							\\
							|\BsrDI|4_{F}|6_{1}|4_{F}|6|
							\BserI|\FokI|5|4_{F}|12|\BpmII|\BpmID|8|6_{0}|8|\BbvI|
						\end{gathered}
					\end{equation}
					
  					After enzymes BsrDI and BbvI have cleaved the molecule, we obtain:
					\begin{equation}
						\begin{gathered}
							\left(
							6_{\beta}|4_{F}|6_{\beta}|\frac{\scriptscriptstyle F_{1}F_{2}}{}|
							~~~~~~~~~~~~~~~~~~~~~~~~~~~~~~~~~~~~~~~~~~~~~~~~~~~~~~~~~~~~~~~~~~~~~~~~~~~
							|\frac{\scriptscriptstyle A_{3}
							A_{4}A_{5}A_{6}}{}|\cdots
							\right)
							\\
							|\frac{\scriptscriptstyle ~~~~~F_{3}F_{4}}{\scriptscriptstyle
							\widetilde{F_{1}}\widetilde{F_{2}}\widetilde{F_{3}}\widetilde{F_{4}}}|6_{1}|4_{F}|6|
							\BserI|\FokI|5|4_{F}|12|\BpmII|\BpmID|8|\frac{\scriptscriptstyle A_{1}A_{2}~~~~~~~~~~~}{\scriptscriptstyle
							\widetilde{A_{1}}\widetilde{A_{2}}\widetilde{A_{3}}\widetilde{A_{4}}\widetilde{A_{5}}\widetilde{A_{6}}}|
 						\end{gathered}
					\end{equation}
					
					Here, the sticky-ends are complementary, which means that they are susceptible of binding together. After this happens
					we obtain:
					\begin{equation}
						\left(
						6_{\beta}|4_{F}|6_{\beta}|4_{F}|6_{1}|4_{F}|6|
						\BserI|\FokI|5|4_{F}|12|\BpmII|\BpmID|8|6_{0}|4_{F}|6_{\lambda_{3}}|4_{F}|\cdots
						\right)
					\end{equation}
					
					Next, enzyme BpmI acts upon the string, which becomes:
					\begin{equation}
						\left(
						6_{\beta}|4_{F}|6_{\beta}|4_{F}|6_{1}|4_{F}|6|
						\BserI|\FokI|5|\frac{\scriptscriptstyle F_{1}F_{2}}{}|~|\frac{\scriptscriptstyle ~~~~~F_{3}F_{4}}{\scriptscriptstyle
						\widetilde{F_{1}}\widetilde{F_{2}}\widetilde{F_{3}}\widetilde{F_{4}}}|6_{\lambda_{3}}|4_{F}|\cdots
						\right)
					\end{equation}
					
					Which becomes the following strand after the sticky-ends bind together:
					\begin{equation}
						\label{teo:paso55}
						\left(
						6_{\beta}|4_{F}|6_{\beta}|4_{F}|6_{1}|4_{F}|6|
						\BserI|\FokI|5|4_{F}|6_{\lambda_{3}}|4_{F}|\cdots
						\right)
					\end{equation}
					
					This last strand can be considered as a new initial configuration for the machine. We develop this branch of the computation
					below. Meanwhile, it is enough to say that currently the head is placed over character $\lambda_{3}$, and it has just written
					the character $1$ on the tape.
				\item[CASO 3.3: $<S_{2}, \lambda_{2}=1>$] Then, the string (\ref{teo:paso43}) is actually the string:

					\begin{equation}
						\left(
						6_{\beta}|4_{F}|6_{\beta}|\frac{\scriptscriptstyle F_{1}F_{2}}{}|~|\frac{\scriptscriptstyle B_{3}
						B_{4}B_{5}B_{6}}{}|4_{F}|
						6_{\lambda_{3}}|4_{F}|\cdots
						\right)
					\end{equation}
					
					For this combination of input and state we expect transition $T_{8}$. When the molecule (\ref{mol:T8}) representing this
					transition attempts to bind to the main molecule, we see:
					\begin{equation}
						\begin{gathered}
							\left(
							6_{\beta}|4_{F}|6_{\beta}|\frac{\scriptscriptstyle F_{1}F_{2}}{}|
							~~~~~~~~~~~~~~~~~~~~~~~~~~~~~~~~~~~~~~~~~~~~~~~~~~~~~~~~~~~~~~~~~~~~~~
							|\frac{\scriptscriptstyle B_{3}
							B_{4}B_{5}B_{6}}{}|\cdots
							\right)
							\\
							|\BsrDI|4_{F}|6_{0}|4_{F}|6|
							\BserI|\FokI|5|4_{F}|12|\BpmII|\BpmID|8|6_{1}|8|\BbvI|
						\end{gathered}
					\end{equation}
					
					This molecule is not yet able to bind to the main molecule. It is only when enzymes BsrDI and BbvI recognise their sites that
					the transition molecule exhibits sticky-ends that are complementary to the main molecule:
					\begin{equation}
						\begin{gathered}
							\left(
							6_{\beta}|4_{F}|6_{\beta}|\frac{\scriptscriptstyle F_{1}F_{2}}{}|
							~~~~~~~~~~~~~~~~~~~~~~~~~~~~~~~~~~~~~~~~~~~~~~~~~~~~~~~~~~~~~~~~~~~~~~~~~~~
							|\frac{\scriptscriptstyle B_{3}
							B_{4}B_{5}B_{6}}{}|\cdots
							\right)
							\\
							|\frac{\scriptscriptstyle ~~~~~F_{3}F_{4}}{\scriptscriptstyle
							\widetilde{F_{1}}\widetilde{F_{2}}\widetilde{F_{3}}\widetilde{F_{4}}}|6_{0}|4_{F}|6|
							\BserI|\FokI|5|4_{F}|12|\BpmII|\BpmID|8|\frac{\scriptscriptstyle B_{1}B_{2}~~~~~~~~~~~}{\scriptscriptstyle
							\widetilde{B_{1}}\widetilde{B_{2}}\widetilde{B_{3}}\widetilde{B_{4}}\widetilde{B_{5}}\widetilde{B_{6}}}|
						\end{gathered}
					\end{equation}
					
					When the complementary sticky-ends bing together, they yield the following molecule:
					\begin{equation}
						\left(
						6_{\beta}|4_{F}|6_{\beta}|4_{F}|6_{0}|4_{F}|6|
						\BserI|\FokI|5|4_{F}|12|\BpmII|\BpmID|8|6_{1}|4_{F}|6_{\lambda_{3}}|4_{F}|\cdots
						\right)
					\end{equation}
					
					Next, the enzyme BpmI cleaves the molecule after recognising its site. This results in:
					\begin{equation}
						\left(
						6_{\beta}|4_{F}|6_{\beta}|4_{F}|6_{0}|4_{F}|6|
						\BserI|\FokI|5|\frac{\scriptscriptstyle F_{1}F_{2}}{}|~|\frac{\scriptscriptstyle ~~~~~F_{3}F_{4}}{\scriptscriptstyle
						\widetilde{F_{1}}\widetilde{F_{2}}\widetilde{F_{3}}\widetilde{F_{4}}}|6_{\lambda_{3}}|4_{F}|\cdots
						\right)
					\end{equation}
					
					After the complementary sticky-ends bind together, we obtain the following molecule:
					\begin{equation}
						\label{teo:paso61}
						\left(
						6_{\beta}|4_{F}|6_{\beta}|4_{F}|6_{0}|4_{F}|6|
						\BserI|\FokI|5|4_{F}|6_{\lambda_{3}}|4_{F}|\cdots
						\right)
					\end{equation}
					
					This strand can be considered a new initial configuration. Moreover, we know that the head is currently placed over the 
					character $\lambda_{3}$ and that the character $0$ has just been written recently on the tape.
 			 \end{description}
			 
			At this point, it is natural to wonder what happens if we further develop the \textbf{CASES} recently presented. In the following observation
			we give an answer to this question. 
			
			\begin{Obs}
				Let us take a look at the tapes that result from applying \textbf{CASES} \textbf{3.1}, \textbf{3.2} y \textbf{3.3}, that is, strings
				(\ref{teo:paso49}), (\ref{teo:paso55}) and (\ref{teo:paso61}). As noted previously, these strings are similar to each other and to
				the initial configuration (\ref{teo:conf_ini}). Thus, we can expect the same behaviour from our TM for the rest of the computation.
				In other words, we have exhausted all the possible behaviours of the machine, thus its operation will ever fall in one of the cases
				described above.
			\end{Obs}
			
			The above not only proves that our transition molecules work correctly on the molecular tape, but also that they write what is expected
			on the tape.\qed			
		\end{proof}
	
\end{document}